\documentclass[prb,a4paper,aps,10pt,twocolumn,floatfix,superscriptaddress,amsfonts]{revtex4-2}
\pdfoutput=1
\usepackage[T1]{fontenc}\usepackage[latin1]{inputenc}
\usepackage{dcolumn,graphicx,color,booktabs,microtype,afterpage} 
\usepackage{sidecap}
\usepackage{times}
\usepackage[colorlinks,plainpages=false,linkcolor=blue,urlcolor=blue,citecolor=blue,pdfpagemode=UseNone,pdfstartview=FitBH]{hyperref}
\usepackage{mhchem}
\usepackage{amssymb}
\usepackage{amsmath}
\usepackage{graphicx}
\usepackage{esvect}
\usepackage{bm}
\usepackage{tabularx}
\usepackage{subfigure}
\begin{document}
%
%
%
%
\title{Magnetic Phase Diagram of ErB$_4$ as Explored by Neutron Scattering}

\author{Simon Flury}
\email{simon.flury@psi.ch}
\affiliation{Physik-Institut, Universit{\"a}t Zurich, Winterthurerstrasse 190, CH-8057 Zurich, Switzerland.}
\affiliation{Laboratory for Neutron and Muon Instrumentation, PSI Center for Neutron and Muon Sciences, CH-5232 Villigen, Switzerland.}
\author{Wolfgang J. Simeth}
\affiliation{Physik-Institut, Universit{\"a}t Zurich, Winterthurerstrasse 190, CH-8057 Zurich, Switzerland.}
\affiliation{Laboratory for Neutron and Muon Instrumentation, PSI Center for Neutron and Muon Sciences, CH-5232 Villigen, Switzerland.}
\author{Danielle R. Yahne}
\affiliation{Physik-Institut, Universit{\"a}t Zurich, Winterthurerstrasse 190, CH-8057 Zurich, Switzerland.}
\affiliation{Laboratory for Neutron and Muon Instrumentation, PSI Center for Neutron and Muon Sciences, CH-5232 Villigen, Switzerland.}
\author{Daniel G. Mazzone}
\affiliation{Laboratory for Neutron Scattering and Imaging,  PSI Center for Neutron and Muon Sciences, CH-5232 Villigen, Switzerland.}
\author{Eric D. Bauer}
\affiliation{$^3$Los Alamos National Laboratory, Los Alamos, NM 87545, USA}
\author{Priscila F. S. Rosa}
\affiliation{$^3$Los Alamos National Laboratory, Los Alamos, NM 87545, USA}
\author{Romain Sibille}
\affiliation{Laboratory for Neutron Scattering and Imaging,  PSI Center for Neutron and Muon Sciences, CH-5232 Villigen, Switzerland.}
\author{Oksana Zaharko}
\affiliation{Laboratory for Neutron Scattering and Imaging,  PSI Center for Neutron and Muon Sciences, CH-5232 Villigen, Switzerland.}
\author{Dariusz J. Gawryluk}
\affiliation{Laboratory for Multiscale Material Experiments,
 PSI Center for Neutron and Muon Sciences, CH-5232 Villigen, Switzerland.}
\author{Marc Janoschek}
\email{marc.janoschek@psi.ch}
\affiliation{Physik-Institut, Universit{\"a}t Zurich, Winterthurerstrasse 190, CH-8057 Zurich, Switzerland.}
\affiliation{Laboratory for Neutron and Muon Instrumentation,  PSI Center for Neutron and Muon Sciences, CH-5232 Villigen, Switzerland.}

\date{\today}
%
%
%
%
\begin{abstract}
The tetragonal $4f$-electron intermetallic ErB$_4$ is characterized by strong Ising anisotropy along the tetragonal $c$ axis. The magnetic moments on the erbium sites can be mapped onto a Shastry-Sutherland (SSL) lattice resulting in geometrical frustration. At zero magnetic field ErB$_4$ exhibits collinear columnar antiferromagnetic (CAFM) order below $T_\text{N} = 15.4$~K. In the presence of a magnetic field parallel to the $c$ axis, ErB$_4$ exhibits a plateau at $1/2$ of the saturation magnetization $M_\text{S}$, which arises at a spin flip transition at $H_1$~$=$~1.9~T. Fractional magnetization plateaus and other exotic spin phases are a well-established characteristic feature of frustrated spin systems. Monte Carlo simulations propose that ErB$_4$ is an ideal candidate to feature a spin supersolid phase in close vicinity of $H_1$ between the CAFM and $M/M_\text{S}=1/2$ plateau (HP) phase. Here we combine single-crystal neutron diffraction and inelastic neutron scattering to study the magnetic phase diagram and the crystal electric field (CEF) ground state of ErB$_4$. Our measurements as a function of magnetic field find no signature of the spin supersolid phase but allow us to determine the magnetic structure of the HP phase to be of the uuud type consistent with an Ising material. The magnetic moment $\mu_{\mathrm{CEF}}$~$=$~$8.96~\mu_B$ expected from the CEF configuration determined by our inelastic neutron scattering measurements is also consistent with the ordered moment observed in neutron diffraction showing that the moments are fully ordered and close to  the Er$^{3+}$ free ion moment (9.6~$\mu_B$).

\end{abstract}
\pacs{xx.xx.mm}
\maketitle
%
%
%

\section{Introduction}

Magnetic frustration generally arises when the sum of all constraints determined by the underlying microscopic magnetic short-range interactions cannot be simultaneously satisfied. This typically results in the suppression of long-range magnetic order and a large degeneracy of the ground state~\cite{yoshida_theory_1996, Diep_frustration, lacroix_frustrated_2010}. In insulating quantum magnets that exhibit well-localized magnetic moments, magnetic frustration is an extensively studied phenomenon resulting in rich and exotic physics ranging from spin-ice~\cite{ramirez_Ice, harris_Ice} to spin-liquids \cite{QSL}, and experimental realizations of Kitaev magnets, which have the potential to host Majorana fermions \cite{Kitaev}. In turn, insulating frustrated magnets are relevant both for our fundamental understanding of solids as well as for applications in quantum computing and quantum information systems~\cite{Kitaev_quantum_mat}. 

In stark contrast, much less attention has been paid to magnetic frustration in metallic quantum materials. This is because the effects of magnetic frustration in metallic systems are substantially weaker due to the long-range nature of itinerant magnetic exchange interactions~\cite{lacroix_frustrated_2010, Julian_Kee}. Nevertheless, it has been recently demonstrated that frustration is also a relevant ingredient for understanding the magnetism in strongly correlated electron materials. Examples are heavy fermion materials~\cite{si_heavy_2010, coleman_frustration_2010, das_magnitude_2014, fobes_low_2017} where metallic quantum spin liquid states have been proposed~\cite{Nakatsuji_PRL_2006, Tokiwa2014}, skyrmion lattice materials \cite{kurumaji_skyrmion_2019, hirschberger_skyrmion_2019, takagi_square_2022}, cubic geometrically frustrated rare-earth intermetallics~\cite{nakamura_partially_1999,stockert_magnetic_2020}, as well as van-der-Waals metals~\cite{okuma_magnetic_2021}. A common theme in all of these metals is the interplay between frustration and magnetic anisotropy.

\begin{figure*}[t!]
\centering
\includegraphics[width=0.7\linewidth]{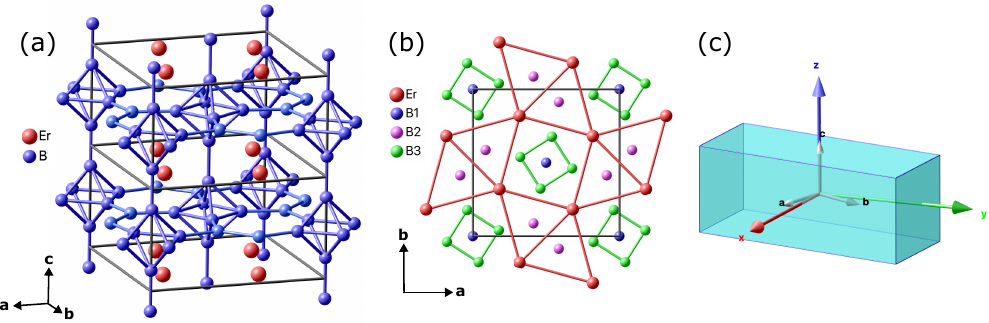}
\caption{(a) The tetragonal crystal structure of ErB$_4$ (space-group \textit{P4/mbm}) is shown. (b) Projection of the crystal structure onto the tetragonal basal plane. The red tubes denote the bonds between the Er atoms, which sit on a Shastry-Sutherland lattice (SSL). The black lines denote the boundaries of the conventional unit cell. (c) The model of the sample shape, which was used to calculate the scattering path lengths for
absorption and extinction correction, is illustrated in the experimental scattering geometry. Here \textbf{a}, \textbf{b} and \textbf{c} define the crystallographic axes, whereas \textbf{x}, \textbf{y} and \textbf{z} represent the laboratory coordinate
system.}
\label{fig:Fig1}
\end{figure*}

Here we study the material ErB$_4$, which belongs to the family of metallic rare-earth tetra-borides $R$B$_4$ ($R$ = rare-earth). The tetraborides crystallizes in the tetragonal space-group \textit{P4/mbm} as illustrated in Fig.~\ref{fig:Fig1}(a). The magnetic moments on the rare-earth sites can be mapped onto a Shastry-Sutherland lattice (SSL)\cite{group} (see Fig. ~\ref{fig:Fig1}(b)). The SSL is a square lattice with antiferromagnetic nearest-neighbour interactions in every square ($J_1$, cf. Fig.~\ref{fig:Fig2}(a)) and antiferromagnetic next-nearest-neighbour interactions in every second square ($J_2$, cf. Fig.~\ref{fig:Fig2}(a)), resulting in geometrical frustration~\cite{SSL_lat}. An iconic realization of the SSL in an insulator is SrCu$_2$(BO$_3$)$_2$ \cite{SrCu2(BO3)2}, which at low temperature exhibits a sequence of magnetization plateaus at fractional values of the saturation magnetization $M_s$, and a plethora of exotic properties and excitations~\cite{shi_discovery_2022, nojiri_study_2000}. Interestingly, despite being metals, various members of the rare-earth tetra-borides similarly exhibit magnetization plateaus at fractional values~\cite{DyB4,Pfeiffer,eTransport,HoB4,Navid,Yoshii,Michimura,Siemensmeyer} demonstrating that frustration is an important tuning parameter. Here the magnetic structure and the number of successive magnetic transitions as a function of magnetic field resulting in the plateau phases changes depending on the rare-earth ion. HoB$_4$\cite{HoB4} and TmB$_4$\cite{Michimura,Siemensmeyer} feature complex magnetic order with plateaus at different fractions of $M_\text{S}$. Even rare-earth tetra-borides with seemingly simple magnetic structures can show plateau phases. TbB$_4$\cite{Navid,Yoshii} orders in an antiferromagnetic structure, with magnetic moments within the $a$-$b$ plane, and shows multiple plateaus at $M/M_\text{S} = 2/9,1/3,4/9,1/2,...$ under applied magnetic field along the c-axis.

ErB$_4$ is well suited to investigate the interplay of magnetic frustration and anisotropy due to the Ising-like character of its magnetism \cite{Pfeiffer,eTransport,Schaefer}. In fact, several theoretical studies have treated the SSL in the strong Ising limit and have identified ErB$_4$ as an ideal candidate to experimentally study this problem
~\cite{dublenych,huo,moliner,Frustration_ErB4,supersolid}. ErB$_4$ orders in a collinear columnar antiferromagnetic (CAFM) ground state below $T_\text{N} = 15.4$~K where the Er moments point along the crystallographic c-axis (cf. Fig.~\ref{fig:Fig2}(a))~\cite{eTransport, Schaefer}. Neutron diffraction studies~\cite{Pfeiffer,Schaefer} and magnetization~\cite{eTransport} measurements for magnetic fields applied along the tetragonal $c$ axis further revealed a meta-magnetic transition to a ferrimagnetic phase with $\text{M/M}_{S} = 1/2$ at $H_\mathrm{1}$~$=$~1.9~T (see Figs.~\ref{fig:Fig2}(b) and (c)). However, the full magnetic structure of the $1/2 M_\text{S}$-plateau phase (HP) above $H_\mathrm{1}$ has previously not been determined.

Based on extended Monte Carlo simulations, it has further been proposed that ErB$_4$ may host exotic spin phases in magnetic field~\cite{supersolid}. Notably, a supersolid phase is expected to exist in a narrow region between the CAFM and HP phase. Similarly, one expects a spin superfluid in the vicinity of the phase transition between the HP phase and the field polarized (FP) state. However, stabilizing the spin superliquid and -solid phases requires the existence of additional higher-order exchange interactions $J_3$ and $J_4$, which are illustrated in Fig.~\ref{fig:Fig2}(a). Here $J_3$ is ferromagnetic and $J_4$ is much weaker and antiferromagnetic. Because for ErB$_4$ the local magnetic moments arising from $4f$-electrons on the erbium sites are embedded into a metallic host, long-range Ruderman-Kittel-Kasuya-Yosida (RKKY) interactions mediated by the surrounding conduction electrons may indeed lead to finite exchange integrals $J_3$ and $J_4$. The authors of Ref. \onlinecite{supersolid} further argue based on their Monte Carlo simulations that the sequence of phases observed in ErB$_4$ may only be stabilized in case $J_3$ and $J_4$ are finite.

In this report, we present single-crystal neutron diffraction studies of the magnetic phase diagram of ErB$_4$ in order to investigate whether ErB$_4$ shows signatures in the neutron scattering data that agree with the calculated magnetic structure factor for the supersolid phase~\cite{supersolid}. In addition, we present an inelastic neutron scattering study of the crystal electrical field ground state. 

\section{Experimental Methods}

For our neutron diffraction study, a single crystal of rectangular shape with dimensions of roughly $3\times 1.2\times 1.2$ $\text{mm}^3$ was used. The single crystal was grown using the traveling solvent floating zone method \cite{behr_floating_zone} starting from high-purity starting materials. To reduce neutron absorption the sample was enriched up to $99.78$ \% by $^{11}$B. Additional $^{11}$B enriched high-quality single crystals were grown using the flux method with Al-flux~\cite{RosaFisk}. Two of these crystals  with a total mass of $m = 0.105$ g where stacked in a mosaic for inelastic neutron scattering experiments with a combined mosaic spread of less than 0.5~degrees.

The neutron diffraction experiment was carried out on the thermal single crystal diffractometer ZEBRA at Paul Scherrer Institute (PSI). The neutron wavelength $\lambda$~$=$~$1.178$~\r{A} as provided by the $(311)$ reflection of a Ge-monochromator was used. The crystal was aligned with the crystallographic $c$ axis vertically. A magnetic field up to $6$~T along the $c$ axis was provided by a vertical field cryomagnet that allows to perform diffraction experiments down to $1.8$ K. The nuclear- and magnetic structure refinements were carried out using the program MAG2POL \cite{mag2pol}. Note that even though our sample was enriched with $^{11}$B (absorption cross section 0.0055 barn) the absorption of $^{10}$B is so large (3835.(9) barn) that even a small fraction results in strong absorption effects and poses a challenge with regard to high-quality structural refinements. In order to overcome this challenge, we determined the transmission factor and applied an absorption correction by using a convex-hull sample shape model in Mag2Pol, which enables the calculation of beam path lengths inside the sample, before and after the scattering process. The model used for the sample shape is illustrated in Fig.~\ref{fig:Fig1}(c). By applying the \textit{on-the-fly} absorption option~\cite{mag2pol} of MAG2POL, the linear absorption coefficient, which highly depends on the $^{11}$B concentration, is recalculated in every iteration of the nuclear refinement using the $^{11}$B concentration as a refinable parameter. 

We further carried out inelastic neutron scattering measurements on the multiplexing spectrometer CAMEA~\cite{lass_commissioning_2023,lass_mjolnir_2020} at PSI to determine the crystal electrical field (CEF) excitation of ErB$_4$. Various incident energies between $E_i = 5$ and $13$~meV were used to investigate the excitation spectrum over an energy transfer of $0$-$10$~meV and a $Q$ transfer $\sim$$0.5$-$2$~\AA$^{-1}$, at $1.6$~K. Preliminary inelastic neutron scattering measurements were additionally performed on the time-of-flight spectrometer FOCUS~\cite{jansen_focus_1997} located at PSI. The neutron data were analyzed using the python package \textit{PyCrystalField}~\cite{scheie_pycrystalfield_2021}.

\section{Experimental Results}
\subsection{Refinement of Crystal Structure}

\begin{figure}[t]
\centering
\includegraphics[width=0.7\linewidth]{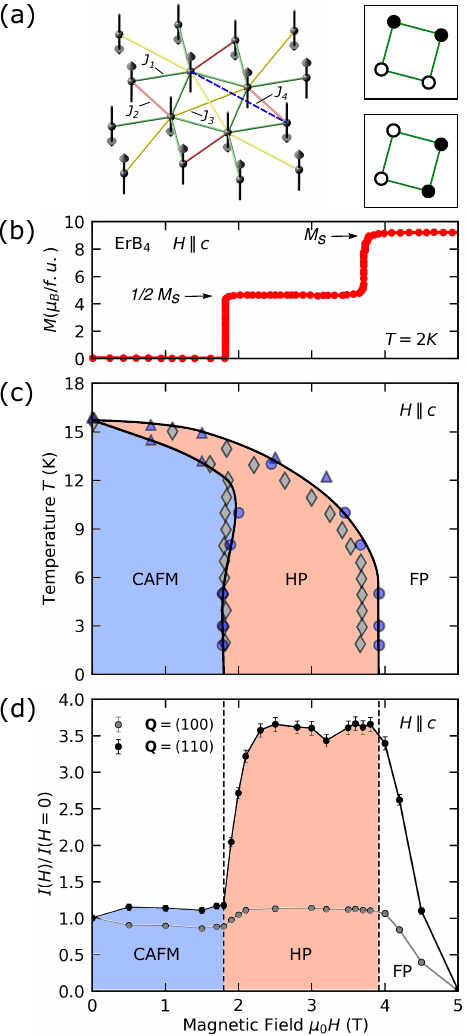}
\caption{(a) The magnetic structure of ErB$_4$ at zero magnetic field is illustrated. The diagonal bond  $J_1$ (green) and the square bond $J_2$ (red) according to the Shastry-Sutherland model are also indicated . For ErB$_4$ the additional exchange interactions to higher order neighbors $J_3$ and $J_4$ are also relevant and are shown as yellow solid and blue dashed lines, respectively (see text for details). The top and bottom insets on the right show the two domains \textit{Pb'am} and \textit{Pba'm} of the magnetic symmetry group \textit{Pbam}, where up-moments are depicted by black filled circles and down-moments depicted by white filled circles. The magnetization as a function of magnetic field applied along the c-axis, measured at $T=2~\text{K}$ (data adapted from Ref.~\cite{eTransport}) and the corresponding temperature $T$ vs. magnetic field $H$ phase diagram derived from transport measurements (grey diamond symbols, from Ref.~\cite{eTransport}) are shown in (b) and (c) respectively. The blue triangles and circles denote phase transitions as determined from our neutron diffraction data using the magnetic field and temperature sweeps shown in Figs.~\ref{fig:Fig4} and ~\ref{fig:Fig5}, respectively. (d) Magnetic field dependence of the relative intensities for different Bragg peak positions measured at $T$~$=$~1.8~K. The plotted intensity of each reflection was normalized with regard to the intensity observed at zero magnetic field.}
\label{fig:Fig2}
\end{figure}

\begin{figure}[ht]
\centering
\vspace*{0.2cm}
\includegraphics[width=0.65\linewidth]{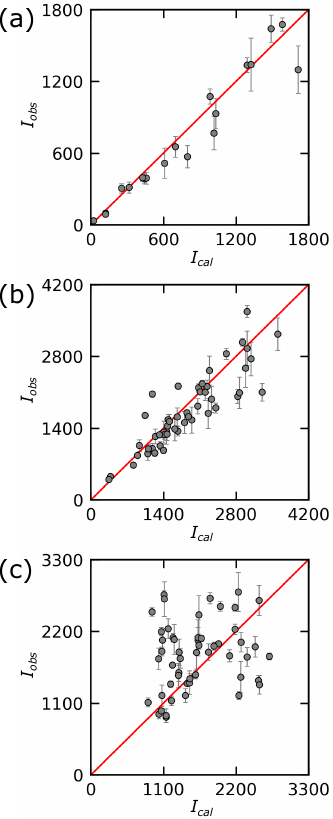}
\caption{Calculated integrated intensities ($I_{cal}$) versus measured integrated intensities ($I_{obs}$) for the best fits obtained in our refinements of all measured Bragg peaks. The red line indicates the ideal condition of $I_{obs} = I_{cal}$. Panels (a), (b), and (c) show the refinements for the nuclear structure measured at $T$~$=$~20~K and $H$~$=$~0, the zero-field CAFM magnetic structure at $T$~$=$~1.8~K and $H$~$=$~0 and the HP phase at $T$~$=$~1.8~K and $H$~$=$~3~T, respectively.}
\label{fig:Fig3}
\end{figure}

To confirm the crystal structure of our samples, the nuclear structure factor was measured on $112$ structural Bragg reflections indexed in the space-group \textit{P4/mbm} (No. 127, see Figure ~\ref{fig:Fig1}(a)) in the paramagnetic state at $20$ K. The structural refinement results in an agreement factor of $R_F = 6.88$ for the previously reported tetragonal crystal structure of ErB$_4$. The calculated versus the observed integrated intensities for the best fit are shown in Fig.~\ref{fig:Fig3}(a). The refined parameters were the atomic positions, two of the diagonal elements $(x_{11},x_{22})$ of the extinction correction tensor and an overall scale factor. The third diagonal element $x_{33}$ could not be refined due to the small number of accessible out of plane reflections $(hkl), l \neq 0$. Due to the relatively small number of recorded reflections combined with a neutron absorbing sample, refinement of the isotropic temperature factor was not possible and therefore fixed to zero. Finally, we also refined the site occupation boron sites with $^{11}$B and obtained $0.987(6)$, which is in good agreement with the purity of $^{11}$B enriched boron used for the growth. The original data set was corrected for absorption and averaged in \textit{P4/mbm} symmetry using the refined $^{11}$B occupancy. The refined atomic positions are tabulated in Table~\ref{table:nuclear} together with the  extinction parameters $x_{11} = 1.09(4)$ and $x_{22} = 0.86(3)$. Our results agree well with previous structural refinements of ErB$_4$ ~\cite{will_neutron_1979,will_neutron_1981}.

\begin{table}[b]
\vspace*{-0.4cm}
\caption{Atomic parameters for the tetragonal crystal structure (space-group \textit{P4/mbm}) of ErB$_4$ as obtained via refinement of our neutron diffraction data. The table lists all atoms, their
Wyckoff-positions and their coordinates in a conventional tetragonal unit cell.}
\label{table:nuclear}
\begin{tabular}{c|c|c|c|c}
    \hline
     Atom & Wyckoff & x & y & z  \\
     \hline
     Er & 4g & 0.3193(7) & 0.8193(7) & 0.0000  \\
     \hline
     B(1) & 4e & 0.0000 & 0.0000 & 0.2045(3)  \\
     \hline
     B(2) & 4g & 0.0849(1) & 0.5849(1) & 0.5000  \\
     \hline
     B(3) & 8j & 0.1767(9) & 0.0400(9) & 0.5000  \\
     \hline
     \hline
\end{tabular}
\vspace*{-0.2cm}
\end{table}

\subsection{Crystal Electric Field Excitations}

\begin{figure}[th!]
\centering
\includegraphics[scale=1]{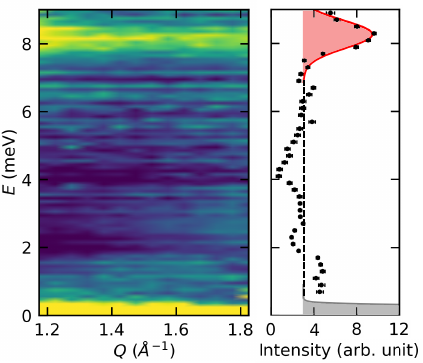}
\caption{Inelastic neutron-scattering measurement up to an energy loss of $9.2$ meV at $1.6$ K. The energy vs intensity plot on the right is the $Q$-integrated, where the red line shows the fit of the CEF-level using the CEF Hamiltonian in Eq.~\ref{equation:CEF_Ham_ErB4}. The black dashed line illustrates the background level and the gray line the peak shape of the elastic line.}
\label{fig:CEF}
\end{figure}

The magnetic moment on the erbium sites in ErB$_4$ is a result of their partially filled 4$f$ shell. The electronic configuration for the Er$^{3+}$ ions is [Xe]4f$^{11}$ and, following Hund's rules, the total angular momentum of the spin-orbit ground state is $J$ = 15/2 with $L$ = 6 and $S$ = 3/2. In absence of the crystal electric field, the corresponding spin-orbit ground state is 2$J$ + 1 = 16-fold degenerate. The neighboring B ions generate the CEF, which typically corresponds to an energy scale of about 100 meV. Note that for Er$^{3+}$ the spin-orbit gap between the first-excited and the spin-orbit ground state is $\lambda J$~$\approx$~2.5 eV\cite{blume_theory_1964}. In turn, for ErB$_4$, we can limit ourselves to lower-$J$ multiplets for our CEF analysis. 

The form of the CEF Hamiltonian is defined by the position of the boron ligands surrounding the magnetic Er$^{3+}$ ion. Using \textit{PyCrystalField}~\cite{scheie_pycrystalfield_2021}, we built a point charge model containing 18 ligands and a mirror plane along $[1\overline{1}0]$ direction. The mirror symmetry suppresses $-m$ terms such that the CEF Hamiltonian is given by:
\begin{eqnarray}
    \mathcal{H}_{CF}^{\text{ErB}_4} &=& B_2^0\mathrm{O}_2^0 +B_2^2\mathrm{O}_2^2 +B_4^0\mathrm{O}_4^0 +B_4^2\mathrm{O}_4^2 +B_4^4\mathrm{O}_4^4 + \nonumber\\
    &+& B_6^0\mathrm{O}_6^0 + B_6^2\mathrm{O}_6^2 +B_6^4\mathrm{O}_6^4 +B_6^6\mathrm{O}_6^6,
    \label{equation:CEF_Ham_D_4h}
\end{eqnarray}
where the $O^{m}_{n}$ are Stevenson operators, which are written in terms of the $J^{+}$, $J^{-}$, and $J_z$ operators~\cite{stevens_matrix_1952}. The calculated CEF parameters of the underlying point charge model reveal that Eq.~\ref{equation:CEF_Ham_D_4h} is predominantly determined by the first two terms allowing us to approximate the CEF Hamiltonian for ErB$_4$ by
\begin{align}
    \mathcal{H}_{CF}^{\text{ErB}_4} \approx B_2^0\mathrm{O}_2^0 +B_2^2\mathrm{O}_2^2.\label{equation:CEF_Ham_ErB4}
\end{align} 

Fig.~\ref{fig:CEF} shows the results of our inelastic neutron measurements on CAMEA. In the left panel of Fig.~\ref{fig:CEF}, we show the neutron energy loss channel up to $9.2$~meV as a function of momentum transfer. Only a single dispersionless excitation is observed at $8.25$~meV in this energy window. As CEF excitations are single-ion properties, they typically lack dispersion and exhibit an intensity distribution, which decreases as a function of momentum transfer following the magnetic form factor of Er$^{3+}$.

Because these measurements were carried out at a temperature $T$~$=$~1.6~K, the first-excited state cannot be significantly thermally populated. Therefore, this excitation stems from the CEF ground state. Notably, of the total of seven transitions between the eight Kramers doublets belonging to the lowest-energy manifold for Er$^{3+}$ with $J = \frac{15}{2}$, this is the first transition between the ground state and the next-highest doublet.

To determine the CEF Hamiltonian, we have fitted the momentum-transfer integrated neutron spectrum shown in the right panel of Fig.~\ref{fig:CEF} to Eq.~\ref{equation:CEF_Ham_ErB4}. The resulting parameters are $B_2^0\mathrm{O}_2^0$~$=$~0.14042 meV and $B_2^2\mathrm{O}_2^2$~$=$~-0.26240 meV, respectively. Tab.~\ref{tab:Eigenvectors} shows the calculated eigenvalues and eigenvectors based on the fitted CEF Hamiltonian up to the first excited doublet. The expected magnetic moment calculated from the fitted CEF-Hamiltonian is $\mu_{\mathrm{CEF}}$~$=$~$8.96(3)~\mu_B$. This compares well with the magnetic saturation moment identified previously in magnetization measurements~\cite{eTransport}.

\begin{table*}[ht]
\centering
\caption{Eigenvalues and eigenvectors for crystal electrical field Hamiltonian in Eq.~\ref{equation:CEF_Ham_ErB4} for the Er$^{3+}$ ions in ErB$_4$ fitted to the inelastic neutron data shown in Fig.~\ref{fig:CEF}. The values are shown up to the first excited doubled of the Er$^{3+}$ $J = \frac{15}{2}$ ground state manifold that was accessible in our measurements. The top numbers in a row correspond to $|+J_z \rangle$ and the bottom numbers to $|-J_z \rangle$.}
\renewcommand{\arraystretch}{1.5}
\begin{tabular}{c|cccccccccccccccc}
\hline\hline
E (meV) &$| \begin{smallmatrix} + \\ - \end{smallmatrix}\frac{15}{2}\rangle$ & $| \begin{smallmatrix} + \\ - \end{smallmatrix}\frac{13}{2}\rangle$ & $| \begin{smallmatrix} + \\ - \end{smallmatrix}\frac{11}{2}\rangle$ & $| \begin{smallmatrix} + \\ - \end{smallmatrix}\frac{9}{2}\rangle$ & $| \begin{smallmatrix} + \\ - \end{smallmatrix}\frac{7}{2}\rangle$ & $| \begin{smallmatrix} + \\ - \end{smallmatrix}\frac{5}{2}\rangle$ & $| \begin{smallmatrix} + \\ - \end{smallmatrix}\frac{3}{2}\rangle$ & $| \begin{smallmatrix} + \\ - \end{smallmatrix}\frac{1}{2}\rangle$  \tabularnewline
\hline
0.000 & $\begin{matrix} -0.0048 \\ 0.0 \end{matrix}$ & $\begin{matrix} 0.0 \\ -0.021 \end{matrix}$ & $\begin{matrix} -0.062 \\ 0.0 \end{matrix}$ & $\begin{matrix} 0.0 \\ -0.1418 \end{matrix}$ & $\begin{matrix} -0.2651 \\ 0.0 \end{matrix}$ & $\begin{matrix} 0.0 \\ -0.4167 \end{matrix}$ & $\begin{matrix} -0.5595 \\ 0.0 \end{matrix}$ & $\begin{matrix} 0.0 \\ -0.647 \end{matrix}$  \tabularnewline 
\hline
0.000 & $\begin{matrix} -0.0048 \\ 0.0 \end{matrix}$ & $\begin{matrix} 0.0 \\ -0.021 \end{matrix}$ & $\begin{matrix} -0.062 \\ 0.0 \end{matrix}$ & $\begin{matrix} 0.0 \\ -0.1418 \end{matrix}$ & $\begin{matrix} -0.2651 \\ 0.0 \end{matrix}$ & $\begin{matrix} 0.0 \\ -0.4167 \end{matrix}$ & $\begin{matrix} -0.5595 \\ 0.0 \end{matrix}$ & $\begin{matrix} 0.0 \\ -0.647 \end{matrix}$  \tabularnewline 
\hline
8.254 & $\begin{matrix} 0.0 \\ 0.0182 \end{matrix}$ & $\begin{matrix} -0.0709 \\ 0.0 \end{matrix}$ & $\begin{matrix} 0.0 \\ 0.1808 \end{matrix}$ & $\begin{matrix} -0.3442 \\ 0.0 \end{matrix}$ & $\begin{matrix} 0.0 \\ 0.508 \end{matrix}$ & $\begin{matrix} -0.5771 \\ 0.0 \end{matrix}$ & $\begin{matrix} 0.0 \\ 0.4686 \end{matrix}$ & $\begin{matrix} -0.1814 \\ 0.0 \end{matrix}$  \tabularnewline 
\hline
8.254 & $\begin{matrix} -0.0182 \\ 0.0 \end{matrix}$ & $\begin{matrix} 0.0 \\ 0.0709 \end{matrix}$ & $\begin{matrix} -0.1808 \\ 0.0 \end{matrix}$ & $\begin{matrix} 0.0 \\ 0.3442 \end{matrix}$ & $\begin{matrix} -0.508 \\ 0.0 \end{matrix}$ & $\begin{matrix} 0.0 \\ 0.5771 \end{matrix}$ & $\begin{matrix} -0.4686 \\ 0.0 \end{matrix}$ & $\begin{matrix} 0.0 \\ 0.1814 \end{matrix}$  \tabularnewline
\hline \hline
\end{tabular}
\label{tab:Eigenvectors}
\end{table*}

\subsection{Magnetic Phase Diagram probed by Neutron Diffraction}

\begin{table*}[t]
	\caption{Irreducible representation and basis vector composition for space group $P4/mbm$ with $\bm{k} = 0$ found using the representation analysis option of MAG2POL. The Erbium atomic positions are defined as Er1 = (0.31875,0.81875,0), Er2 = (0.68125,0.18125,0), Er3 = (0.18125,0.31875,0), Er4 = (0.81875,0.68125,0).}
	\begin{ruledtabular}
		\begin{tabular}{cccccc | cccccc}
		   IR  &   BV & Atom &   \multicolumn{3}{c}{BV components} & IR  &   BV & Atom &   \multicolumn{3}{c}{BV components} \\
        \toprule
        $\Gamma_2$ & $\Psi_1$ & Er1 &  0.707 & 0.707 & 0 & $\Gamma_9$ & $\Psi_1$ & Er1 &  1 & 0 & 0\\
        (afm)      &          & Er2 &  -0.707 & -0.707 & 0 &   (fm)        &          & Er2 &  1 & 0 & 0\\ 
                   &          & Er3 &  -0.707 & 0.707 & 0 &           &          & Er3 &  0 & -$i$ & 0\\ 
                   &          & Er4 &  0.707 & -0.707 & 0 &           &          & Er4 &  0 & -$i$ & 0\\ 
        \midrule 
        $\Gamma_3$ & $\Psi_1$ & Er1 &  0 & 0 & 1 & $\Gamma_9$ & $\Psi_2$ & Er1 &  0 & 1 & 0\\
         (fm)      &          & Er2 &  0 & 0 & 1 &           &          & Er2 &  0 & 1 & 0\\ 
                   &          & Er3 &  0 & 0 & 1 &           &          & Er3 &  $i$ & 0 & 0\\ 
                   &          & Er4 &  0 & 0 & 1 &           &          & Er4 & $i$ & 0 & 0\\  
        \midrule
        $\Gamma_4$ & $\Psi_1$ & Er1 &  0.707 & -0.707 & 0 & $\Gamma_9$ & $\Psi_3$ & Er1 &  0 & -$i$ & 0\\
        (afm)      &          & Er2 &  -0.707 & 0.707 & 0 &           &          & Er2 &  0 & -$i$ & 0\\ 
                   &          & Er3 &  0.707 & 0.707 & 0 &           &          & Er3 &  -1 & 0 & 0\\ 
                   &          & Er4 &  -0.707 & -0.707 & 0 &           &          & Er4 &  -1 & 0 & 0\\ 
        \midrule
        $\Gamma_5$ & $\Psi_1$ & Er1 &  0 & 0 & 1 & $\Gamma_9$ & $\Psi_4$ & Er1 &  -$i$ & 0 & 0\\
        (afm)      &          & Er2 &  0 & 0 & 1 &           &          & Er2 &  -$i$ & 0 & 0\\ 
                   &          & Er3 &  0 & 0 & -1 &           &          & Er3 &  0 & 1 & 0\\ 
                   &          & Er4 &  0 & 0 & -1 &           &          & Er4 &  0 & 1 & 0\\ 
        \midrule
        $\Gamma_6$ & $\Psi_1$ & Er1 &  0.707 & -0.707 & 0 & $\Gamma_{10}$ & $\Psi_1$ & Er1 &  0 & 0 & 1\\
         (afm)     &          & Er2 &  -0.707 & 0.707 & 0 &    (afm)       &          & Er2 &  0 & 0 & -1\\ 
                   &          & Er3 &  -0.707 & -0.707 & 0 &           &          & Er3 &  0 & 0 & -$i$\\ 
                   &          & Er4 &  0.707 & 0.707 & 0 &           &          & Er4 &  0 & 0 & $i$\\
        \midrule
        $\Gamma_8$ & $\Psi_1$ & Er1 &  0.707 & 0.707 & 0 & $\Gamma_{10}$ & $\Psi_2$ & Er1 &  0 & 0 & $i$\\
         (afm)     &          & Er2 &  -0.707 & -0.707 & 0 &           &          & Er2 &  0 & 0 & -$i$\\ 
                   &          & Er3 &  0.707 & -0.707 & 0 &           &          & Er3 &  0 & 0 & -1\\ 
                   &          & Er4 &  -0.707 & 0.707 & 0 &           &          & Er4 &  0 & 0 & 1\\ 
		\end{tabular}
	\end{ruledtabular}
 \label{table:Irreps}
\end{table*}

Before performing symmetry analysis to determine the magnetic structures from our neutron diffraction data, we first establish the magnetic phase diagram based on neutron scattering data. In Fig.~\ref{fig:Fig2}(d), we show the integrated neutron intensity for the $(100)$ and $(110)$ Bragg reflections. Here the plotted intensity of each reflection was normalized with regard to the intensity observed at zero magnetic field. Due to symmetry, the $(100)$ reflection represents purely antiferromagnetic contributions, whereas the $(110)$ reflection corresponds to an additional uniform ferromagnetic contribution. The relative intensity of the $(110)$ Bragg peak shows a clear plateau as a function of magnetic field, whereas the antiferromagnetic contribution on $(100)$ remains almost unchanged. This shows that an additional uniform ferromagnetic contribution to the magnetization emerges when entering the HP phase, suggesting that the metamagnetic transition above $H_\text{1}$ can be characterized as a spin-flip transition.

\begin{figure}[th]
\centering
\includegraphics[width=0.8\linewidth]{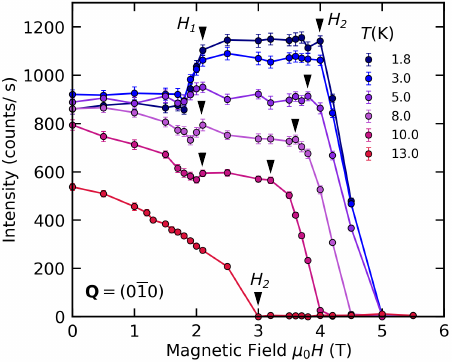}
\caption{Integrated intensity as a function of magnetic field for the purely magnetic Bragg signal at $(0\overline{1}0)$, recorded at different temperatures. The magnetic field was applied along the crystallographic $c$-axis.}
\label{fig:Fig4}
\end{figure}

\begin{figure}[ht]
\centering
\includegraphics[width=0.7\linewidth]{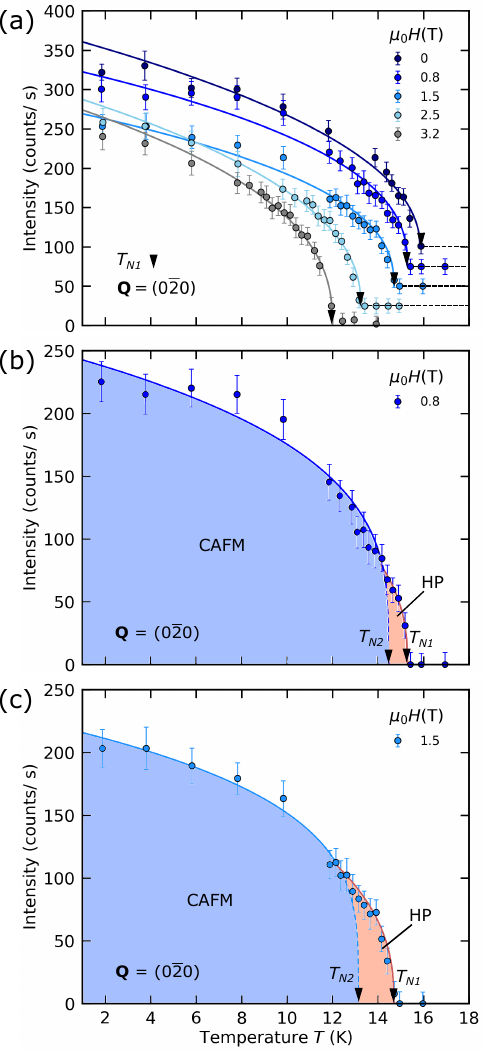}
\caption{(a) Integrated intensity for the $(0\overline{2}0)$ Bragg peak as a function of temperature measured at different magnetic fields. Here we only show the magnetic contribution to the reflection, which was obtained by subtracting the intensity obtained at a temperature well above the critical temperature. For better readability the data sets were shifted by a constant offset of 25 with respect to each other, such that the dashed lines indicate zero intensity for each data sets. The solid lines represent a fit to Eq.~\ref{eq:order parameter}. For intermediate field values $\mu_0 H$~$=$~1.5 and 2.5~T  Eq.~\ref{eq:order parameter} is not suitable to describe the temperature dependence for the entire temperature range below $T_\text{N1}$ (see text for details). As shown in panels (b) and (c) the situation is more complex for $\mu_0 H$~$=$~1.5 and 2.5~T, respectively (see text for details).}
\label{fig:Fig5}
\end{figure}

In Figs.~\ref{fig:Fig4} and ~\ref{fig:Fig5}, we additionally show integrated intensities measured on the $(0\Bar{1}0)$ and $(0\Bar{2}0)$ reflections as a function of temperature $T$ and magnetic field $\mu_0 H$, which we used to establish the $H$-$T$ phase diagram of ErB$_4$. In Fig.~\ref{fig:Fig4}, we illustrate the magnetic field dependence of the integrated intensity of the $(0\overline{1}0)$ pure magnetic Bragg peak for various temperatures $T$. When entering the HP phase the intensity displays a clear jump and stays nearly constant as a function of field analogous to the magnetization plateau. For increasing temperature, the integrated intensity decreases as a function of field and the jump in intensity reduces to a small kink in the data. The field values for the phase transitions between the CAFM and HP phase ($H_\text{1}$) and between the HP and the FP state ($H_\text{2}$) as determined from our neutron data, are denoted with black triangles in Fig.~\ref{fig:Fig4}.

Fig.~\ref{fig:Fig5}(a) shows the temperature dependence of the magnetic contribution to the $(0\overline{2}0)$ Bragg peak for different magnetic fields. Here the magnetic part of the $(0\overline{2}0)$ Bragg peak was obtained by subtracting the Bragg peak signal well above the ordering temperature. To determine the critical temperature $T_\text{N1}$ the integrated intensity $I$ was fitted to  
\begin{equation}
    I(T) = M^2(T) = \left(\frac{T_\text{N1}-T}{T_\text{N1}}\right)^{2\beta}, \label{eq:order parameter}
\end{equation}
where $M$ is the staggered magnetization (order parameter) and $\beta$ is the critical exponent of the order parameter. The best fits are represented by the solid lines in Fig.~\ref{fig:Fig5}(a). The corresponding $T_\text{N1}$ determined by these fits is denoted with a black triangle in Fig.~\ref{fig:Fig5}(a). At zero field, we obtain $\beta = 0.23(1)$ whereas at the highest fields, we find $\beta = 0.20(2)$. However, for intermediate field values $\mu_0 H$~$=$~1.5 and 2.5~T, the temperature dependence (see Fig~\ref{fig:Fig5}(a)) shows a weak shoulder-like feature and Eq.~\ref{eq:order parameter} is not suitable to describe the temperature dependence for the entire temperature range below $T_\text{N1}$. Comparing with the phase diagram previously established from bulk measurements (cf. Ref.~\cite{eTransport}) shown in Fig.~\ref{fig:Fig2}(c), it becomes apparent that for those fields, the system undergoes two phase transitions as a function of temperature. Notably, the shoulder is associated with the phase transition between the CAFM and the HP, which we label with $T_\text{N2}$. 
 
In Figs.~\ref{fig:Fig5}(b) and (c), we demonstrate that the temperature dependence of the integrated intensity $I$ of the $(0\overline{2}0)$ reflection for 1.5 and 2.5~T can be well accounted for by fitting the region above and below $T_\text{N2}$ with distinct critical exponents. Notably, we find $\beta = 0.14(1)$ describes the temperature dependence of the order parameter below $T_\text{N2}$ best.

All phase boundaries that were determined from our neutron diffraction data are also denoted in Fig.~\ref{fig:Fig2}(c). The phase transitions determined from  magnetic field (see Fig.~\ref{fig:Fig4}) and temperature scans (see Fig.~\ref{fig:Fig5}) are marked with blue triangles and circles, respectively. This shows that the phase diagram derived from neutron measurements agrees well with the phase diagram previously constructed from bulk transport measurements in Ref.~\cite{eTransport}.

We note that according to Ref. \onlinecite{supersolid} the magnetic structure factor for the proposed spin supersolid phase should lead to additional magnetic intensities at peaks corresponding to magnetic propagation vectors of the type (100), (110) and (000). The phase is expected in a narrow magnetic field regime around $H_1$. As shown in Figs.~\ref{fig:Fig2}(b) and~\ref{fig:Fig4}, we do not observe any extra intensity in this field range and observed intensities can be attributed either to the CAFM phase or the HP phase as we show below. 

\begin{table}[t]
\centering
\caption{Agreement factors obtained by fitting the irreducible representations to the recorded structure factors at 1.8 K and zero magnetic field. The fits were performed using the software Mag2Pol and the normalized basis vectors of the IRs in Tab.~\ref{table:Irreps}.}
\begin{tabular}[t]{c|c|c|c|c|c|c}
\hline
   IR:  &   $\Gamma_2$ &   $\Gamma_4$ &   $\Gamma_5$ &   $\Gamma_6$ &   $\Gamma_8$ &  $\Gamma_{10}$ \\
\hline
$R_F$: & 21.51  & 20.08 & 30.15 & 21.99 & 20.17  & 7.88\\
\hline
\hline
\end{tabular}
\label{table:fit_Irreps}
\end{table}

\subsection{Magnetic Structure at Zero Magnetic Field}

\begin{figure*}[t]
\centering
\includegraphics[scale=1]{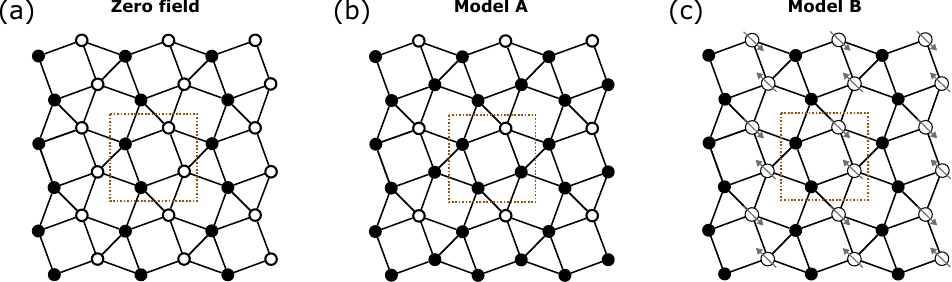}
\caption{(a) Collinear columnar antiferromagnetic (CAFM) structure at zero field. (b,c) Spin-flip models used for the refinement of the magnetic structure in the half-plateau phase. Model A denotes the situation, where starting from the collinear antiferromagnetic groundstate at zero field, one of the moments is flipped $180^\circ$. Model B describes the situation where both down moments are flipped by $90^\circ$, such that they are oriented within the basal plain. These moments are illustrated by gray circles with arrows denoting the moment direction within the plane.} 
\label{models}
\end{figure*}

We now proceed to confirm the magnetic structure at zero magnetic field. The structure is characterized by the propagation vector $\bm{k} = (1,0,0)$\cite{Pfeiffer,Schaefer} that is equivalent to $\bm{k}=0$. By applying group-theoretical symmetry analysis using MAG2POL and SARAh, the decomposition of the representation of the magnetic Er site occupying the 4$g$ Wyckoff-position of the $P4/mbm$ space-group based on $\bm{k}=0$ into irreducible representations (IRs) is given by:
\begin{align}
    \Gamma = \Gamma_2^1 + \Gamma_3^1 + \Gamma_4^1 + \Gamma_5^1 + \Gamma_6^1 + \Gamma_8^1 + 2\Gamma_9^2 + \Gamma_{10}^1,
\end{align}

where the superscripts define the order of the IR. The calculated irreducible representations containing the basis vector components for the different erbium atom positions are presented in Table~\ref{table:Irreps}. The IRs $\Gamma_2$, $\Gamma_4$, $\Gamma_5$, $\Gamma_6$, $\Gamma_8$, $\Gamma_{10}$ are associated with antiferromagnetic configurations and $\Gamma_3$, $\Gamma_{9}$ with ferromagnetic configurations. Due to the vanishing uniform magnetization in the zero-magnetic-field phase (cf. Fig.~\ref{fig:Fig2}(a)), only IRs representing antiferromagnetic configurations are allowed. $152$ Bragg reflections were recorded at a temperature of $1.8$~K to perform the magnetic structure determination. The reflections were integrated and averaged in \textit{Pmmm} symmetry to allow for lower symmetry models. The magnetic form factor was described by the analytic approximation to the $\langle j_0 \rangle$ integrals for the $f$-electrons of the Er$^{3+}$ ions~\cite{formfactor}.

The results of the magnetic refinements are presented in Tab.~\ref{table:fit_Irreps}, showing that the best refinement was obtained for the IR $\Gamma_{10}$, with an agreement factor of $R_\text{F} = 7.88$. Refining the magnetic components $C_1$ and $C_2$ associated with the normalized basis vectors $\Psi^{10}_1, \Psi^{10}_2$ of $\Gamma_{10}$, with the constraint of the magnetic moments on each site having the same modulus, results in $C_1=C_2=8.8(9)~\mu_\text{B}$, with an equal population of the two magnetic domains, 0.50(2) (\textit{Pb'am}), 0.49(2) (\textit{Pba'm}) (cf. Fig.~\ref{fig:Fig2}(a)). The ordered magnetic moment agrees well with the magnetic moment $\mu_{\mathrm{CEF}}$~$=$~$8.96(3)~\mu_B$ expected from the determined crystal-electrical field Hamiltonian. The calculated versus the observed integrated intensities for the best fit to IR $\Gamma_{10}$ are shown in Fig.~\ref{fig:Fig3}(b). The magnetic structure identified in our refinements is in agreement with the zero-field structure proposed in Ref.~\cite{Schaefer} and is illustrated in Figs.~\ref{fig:Fig2}(a) and ~\ref{models}(a).

\subsection{Magnetic Structure of the HP Phase}

As we have demonstrated above, the metamagnetic transition at $H_\text{1}$ is characterized by the emergence of an additional uniform ferromagnetic contribution to the magnetization suggesting a spin-flip transition. Thus, to determine the magnetic structure in the HP phase we study two different spin-flip models that are allowed by the symmetry group \textit{Pbam}, which are illustrated in Figure~\ref{models}. 

Both models have in common that half of the spins in the magnetic unit cell are arranged in pairs, in an antiparallel fashion, such that they do not contribute to the uniform magnetization. In turn, each model results in a uniform magnetization that corresponds to 1/2 of the saturation magnetization $M_S$ so that they comply with the bulk magnetization measurements. Model A represents the Ising-type uuud-model, where one of moments pointing downwards in the zero field ground state flips by $180^\circ$ (cf. Fig.~\ref{models}). Model B represents the situation where both of the spins pointing downwards at $H=0$~T become flipped by $90^\circ$ at $H_1$, such that they are coplanar within the basal plane. 

We note that the refinement of the magnetic structure in the plateau phase is affected by additional challenges. Notably, a first-order phase transition driven by the application of magnetic field, as observed at $H_1$ (cf. jump in the magnetic order shown in Fig.~\ref{fig:Fig2}), typically results in an abrupt rearrangement of magnetic domains. Due to magneto-striction, this also influences the nuclear intensities, which impacts the correction of extinction. In turn, this requires refining the extinction parameters for the plateau phase separately. Thus, the magnetic structure refinements of the HP phase were performed by treating the extinction parameters $\mathrm{x_{11}}$ and $\mathrm{x_{22}}$ as free parameters. 

The separate refinement of $\mathrm{x_{11}}$ and $\mathrm{x_{22}}$, however, posses the problem that a reliable refinement of the size of magnetic moment is difficult. In the case of the ErB$_4$, this can be mitigated, because the ordered moment in the zero-field phase is known and corresponds to the magnetic moment $\mu_{\mathrm{CEF}}$~$=$~$8.96(3)~\mu_B$ expected from the CEF configuration. Moreover, both previous bulk magnetization measurements as well as our neutron diffraction results show that the transition in the HP phase is characterized by a spin-flip transition, which does not affect the magnitude of the magnetic moments. In turn, the refinement was carried out with a fixed magnetic moment $\mu_{\mathrm{CEF}}$ and the scale parameter determined for the zero field magnetic structure. In addition, Model B was refined by choosing the rotation angle $\varphi$ in the basal plane as a refinement parameter under the constraint of a vanishing net magnetic moment within the basal plane. A canting of the moments out of plane did not lead to better refinements. Because, our limited data set is not appropriate to differentiate between models were canting angles differ just by a few degrees, we constrained the moments withing the basal plane. 

The spin-flip model A, where one of the magnetic moments flips, leads to the best agreement factor of $R_F = 14.77$ (see Table.~\ref{table:models}). The refinement of model A is presented in Fig.~\ref{fig:Fig3}(c). The large difference between the calculated and recorded intensities is a direct result of the small number of recorded Bragg peaks, the strong absorption and extinction in the material and the resulting errors propagating through all refinements, but does not change the conclusion that model A provides the best fit of the HP phase in ErB$_4$.

\begin{table}[b]
\caption{Agreement factor and anisotropic extinction parameter of the refined spin-flip models.}
\begin{tabular}{c|c|c|c}
    \hline
    Model & $\mathrm{R_F}$ & $\mathrm{x_{11}}$ & $\mathrm{x_{22}}$   \\
     \hline
     A & 14.77 & 0.10(2) & 0.06(1)\\
     \hline
     B & 17.64 & 0.05(2) & 0.06(2)\\
     \hline
     \hline
\end{tabular}

\label{table:models}
\end{table}


\section{Discussion}

Our neutron scattering study offers insight in the magnetic phase diagram of ErB$_4$. Most notably, our neutron diffraction results in magnetic field show no signatures that may be related to the emergence of a spin supersolid phase between the CAFM and HP phases. This suggests that the exact combination of exchange interactions $J_1$ to $J_4$ that lead to the supersolid phase in the Monte Carlo simulations~\cite{supersolid} does not reflect the magnetic interactions in ErB$_4$. Although, it has been argued in Ref.~\onlinecite{supersolid} that substantial finite ferromagnetic $J_3$ and a small antiferromagnetic $J_4$ are able to stabilize the CAFM and HP phases, we note that various sets of interactions may lead to this sequence of phases. For example, anisotropic interactions $J_2$ in combination with a small $J_3$ may also stabilize these phases~\cite{yadav_observation_2024-1}. However, we also note that our measurements were limited to $T$~$=$~1.8~K and the supersolid phase may arise at lower temperatures. Previous transport and magnetization measurements \cite{eTransport} similarly do not detect any signatures of the supersolid phase. Further, the magnetic phase diagram established via our neutron diffraction experiments agrees well with the previously established phase diagram~\cite{eTransport}. 

Our neutron diffraction results also reveal the critical exponents $\beta$ of the magnetic order parameter for the both the CAFM and the HP phase. For both we find that for the phase transition from the paramagnetic state into the ordered state $\beta$ is approximately 0.22. In contrast for the transition from the the HP into the CAFM phase as a function of temperature we find that $\beta = 0.14(1)$. In principle, due to the strong uniaxial anisotropy of ErB$_4$, we expect that the critical exponents are within the Ising universality class. For two- (2D) and three-dimensional (3D) Ising systems the calculated exponent $\beta$ is 1/8 and 0.326419(3), respectively. This suggests that ErB$_4$ is close to a 2D Ising system. That the critical exponent deviates from 1/8 may be because ErB$_4$ is not a true 2D system, and weak interlayer exchange between different SSL layers exists. This offers an additional explanation of why the supersolid phase is not stabilized in ErB$_4$.

Using symmetry analysis to refine the integrated intensities measured in our single crystal neutron diffraction experiments in applied magnetic field, we further reproduce previous results \cite{Schaefer} and show that the zero-field magnetic ground state is indeed a columnar antiferromagnet. Further for the HP phase, we find that the Ising-like spin-flip model A, corresponding to a uuud structure, provides the best agreement with our data. Although, it is possible that other spin-flip models, where all spins are slightly canted, may be stabilized, our measurements are not sensitive to these small deviations. In addition, it is not expected that such an arrangement of the magnetic moments would be stable over the entire field range of the HP phase, and other plateaus at fractional magnetization values would be expected as observed in TbB$_4$\cite{Navid,Yoshii}. In turn, ErB$_4$ behaves similar to TmB$_4$\cite{Siemensmeyer}, where due to the strong Ising anisotropy, the $1/2M_\text{S}$ plateau arises due to a spin-flip transition in agreement with the uuud-model.

Finally, our inelastic neutron measurements resolve the lowest lying CEF excitation of Er$^{3+}$ with $J = \frac{15}{2}$. This allowed us to determine a minimal CEF Hamiltonian (cf. Eq.~\ref{equation:CEF_Ham_ErB4}). The expected magnetic moment calculated from the fitted CEF-Hamiltonian is $\mu_{\mathrm{CEF}}$~$=$~$8.96(3)~\mu_B$, which agrees well with both the size of the magnetic moments determined from our neutron diffraction results $\mu_{\mathrm{Er}}$~$=$~8.8(9)~$\mu_B$, as well as the saturation magnetization $M_S$~$\approx$~9~$\mu_B$ determined from magnetization measurements~\cite{eTransport}. This demonstrates that the full Er$^{3+}$ moment expected from the CEF ground state orders. Moreover, the ordered moment is close the free ion moment of 9.6~$\mu_B$. This suggest that the magnetic frustration in ErB$_4$ is relatively weak and that, consistent with the observed magnetic phase diagram, its low-temperature magnetic properties are mostly driven by strong Ising anisotropy. This is also consistent with the empirical frustration parameter defined as $f = |\theta_{cw}|/T_N$. Using the Curie-Weiss temperatures $\theta_{CW}= 11.24$~K ($\bm{H}\parallel c$) or $\theta_{CW}= -23.26$~K ($\bm{H}\perp c$) obtained in Ref.~\cite{song_abnormal_2020}, no ($f\sim 0.73$) or very weak ($f\sim 1.5$) magnetic frustration would be expected. 

\section{Summary}
Employing single crystal neutron diffraction and inelastic neutron scattering, we have investigated the magnetic phase diagram and crystal electric field ground state of ErB$_4$. The phase diagram as a function of magnetic field as determined by our study agrees well with previous bulk measurements\cite{eTransport} and shows no sign of the previously proposed spin supersolid phase~\cite{supersolid}. Our measurements confirm the magnetic order at zero magnetic field to be of the CAFM type and show that the spin-flip transition into the HP phase results in a uuud arrangement (cf. Fig.~\ref{models}). The critical exponent $\beta$ of the magnetic order parameter for ErB$_4$ is close to a 2D Ising model, suggesting that ErB$_4$ is not purely characterized by a Shastry-Sutherland model but may exhibit finite interlayer exchange interactions. The crystal electric field ground state of ErB$_4$ determined by our study results in an expected magnetic moment $\mu_{\mathrm{CEF}}$~$=$~$8.96(3)~\mu_B$, showing that the moments in the CAFM and HP phases are fully ordered.


\section{Acknowledgements}

The authors would like to thank to N.Qureshi for giving support using the program MAG2POL. SF acknowledges financial support by the Universit\"at Z\"urich through a UZH Candoc Grant. WSJ and DRY were supported through funding from the European Union's Horizon 2020 research and innovation programme under the Marie Sklodowska-Curie grant agreement No 884104 (PSI-FELLOW-III-3i). DRY and MJ acknowledge funding by the Swiss National Science Foundation through the project ``Berry-Phase Tuning in Heavy f-Electron Metals (\#200650)''. Work at Los Alamos National Laboratory was performed under the U.S. DOE, Office of Science, BES project ``Quantum Fluctuations in Narrow Band Systems''. This work is based on experiments performed at the Swiss spallation neutron source SINQ, Paul Scherrer Institute, Villigen, Switzerland.

\bibliographystyle{apsrev4-1}

\end{document}